\documentclass[a4paper,reprint,aps,pre,floatfix,twocolumn,superscriptaddress,showpacs,nofootinbib]{revtex4-1}
\usepackage{epsfig}
\usepackage{graphics,graphicx,color,subfigure}
\usepackage{amssymb}
\usepackage{amsmath}
\usepackage{amsfonts}
\usepackage[latin1]{inputenc}
\usepackage{comment}
\usepackage{subfigure}
\usepackage{multirow}
\usepackage{color}
\usepackage{hyperref}

\newcommand{\lrangle}[1]{\langle{#1}\rangle}
\newcommand{\lranglec}[1]{\langle{#1}\rangle_{c}}

\newcommand{\vinf}{v_\infty}

\begin{document}
\title{Scaling, cumulant ratios and height distribution of the ballistic 
deposition in 3+1 and 4+1 dimensions}
\author{Sidiney G. Alves}\email{sidiney@ufsj.edu.br}
\affiliation{Departamento de F\'isica e Matem\'atica, Universidade Federal de S\~ao Jo\~ao Del Rei,
36420-000, Ouro Branco, MG, Brazil}
\author{Silvio C. Ferreira}
\email{silviojr@ufv.br}
\affiliation{Departamento de F\'isica, Universidade Federal de Vi\c cosa, 
36570-000, Vi\c cosa, MG, Brazil}

\date{\today}

\begin{abstract}
We investigate the origin of the scaling corrections in ballistic deposition
models in high dimensions using the method proposed by  Alves \textit{et al}.
[Phys Rev. E \textbf{90}, 052405 (20014)] in $d=2+1$ dimensions, where the intrinsic width  associated
with the fluctuations of the height increments during the deposition processes 
is explicitly taken into account. In the present work, we show that this concept holds for 
$d=3+1$ and 4+1 dimensions.  We have found that growth and roughness exponents
and dimensionless cumulant ratios are in agreement with other models, presenting
small finite-time corrections to the scaling, that in principle belong to he
Kardar-Parisi-Zhang (KPZ) universality class in both $d=3+1$ and 4+1. Our
results constitute a new evidence that the upper critical dimension of the
KPZ class, if it exists, is larger than 4.
\end{abstract}

\maketitle

\section{Introduction}

Stochastic growth equations play a central role in the understanding of surface
growth phenomena and are used to classify the different universality classes
\cite{barabasi,meakin}. The Kardar-Parisi-Zhang (KPZ) universality class
introduced by the stochastic equation~\cite{KPZ}
 \begin{equation}
 \frac{\partial h}{\partial t} = \nu \nabla^{2} h + \frac{\lambda}{2}
(\nabla h)^{2} + \xi,
\label{eq:KPZ}
\end{equation}
is one of the most fundamental examples  of nonequilibrium interface growth
model~\cite{krugrev,SasaSpohnJsat,TakeJSP}. Here, $h(\mathbf{x},t)$ represents
the interface height at the position $\mathbf{x}$ and time $t$, the first term
in the right-hand side accounts the relaxation due to the surface tension, the
second one the local lateral growth in the normal direction along the surface
and the last one is a white noise with null mean and amplitude $\sqrt{D}$. The
benchmark of KPZ class is the lateral growth, second term in Eq.~\eqref{eq:KPZ},
that leads to an excess velocity such that the interface envelop moves faster
(or slower if $\lambda<0$) than the rate at which particles are added in the
system.

The interfaces generated by the KPZ equation obey the Family-Vicsek
ansatz~\cite{FV} for the interface width, given by the standard deviation of the
height profile, defined as $w=\sqrt{\lrangle{h^2}-\lrangle{h}^2}$. For a scale of
observation $\ell$ and a growth time $t$, we have that $w(\ell,t)\sim t^{\beta}$
for $t\ll \ell^{\alpha/\beta}$  and $w(\ell,t)\sim \ell^\alpha$ for $t\gg
\ell^{\alpha/\beta}$, where $\alpha$ and $\beta$ are the roughness and growth
exponents, respectively~\cite{barabasi}. The scaling relation $\alpha +
\alpha/\beta =2$, representing Galilean invariance, holds independently of the
dimension~\cite{barabasi}. For $1+1$ dimensions the exponents are exactly known
as $\beta=1/3$ and $\alpha=1/2$~\cite{KPZ}; for higher dimensions exponents are
obtained from simulations~\cite{odor2010,alves14,Kim2014,Marinari2002}. A
thorough analysis  of the KPZ class includes the nature of the underlying
stochastic fluctuations~\cite{TakeJSP,SasaSpohnJsat}. Considering the
non-stationary regime, the height at each surface point evolves as
\begin{equation}
 h = \vinf t + s_\lambda(\Gamma t)^{\beta} \chi+\eta+\ldots,
\label{eq:htcorr}
\end{equation}
where $s_\lambda=\mbox{sgn}(\lambda)$ and $\chi$ is a stochastic variable, whose
distribution is universal and depends on the growth geometries and boundary
conditions~\cite{krug92,PraSpo1,TakeuchiSP,carrasco2014}. The constants $\vinf$
and $\Gamma$ are non-universal and control, respectively, the asymptotic average
velocity  and the amplitude of height fluctuations of the interface. The last
term in the right-hand side of Eq.~\eqref{eq:htcorr} is a non-universal correction
that plays an important role at finite-time analyses in
simulations~\cite{Alves13,Oliveira13R} and experiments
\cite{TakeSano,TakeuchiSP}. It produces a shift in the distribution of the
quantity
\begin{equation}
 q = \frac{h-\vinf t}{s_\lambda(\Gamma t)^{\beta}},
\label{eq:q}
\end{equation}
in relation to the asymptotic  distribution of $\chi$. Except for the very
specific case where $\lrangle{\eta}=0$~\cite{Ferrari}, the 
shift vanishes as $\lrangle{q}-\lrangle{\chi}\sim t^{-\beta}$~\cite{TakeSano,TakeuchiSP,Alves13,Oliveira13R}.
Despite of the absence of exact results in higher dimensions, numerical results
show that the KPZ ansatz  remains valid up to
$d=6+1$~\cite{healyPRL,healyPRE,Oliveira13R,bd_box2d,alves14}.

Discrete growth models are valuable theoretical tools for the realization of
universality classes in surface growth phenomena~\cite{barabasi,meakin} since
they permit to flexibly implement specific physical mechanisms. The ballistic
deposition (BD) model is a paradigmatic interface growth process initially
designed to investigate formation of sediments by the aggregation of small
particles from a colloid dispersion \cite{vold}. In the BD model, particles
move ballistically and normally towards the substrate and are irreversibly
attached at the first contact with the deposit, producing, therefore, lateral
growth that is a central characteristic of the KPZ universality class \cite{KPZ}.
However, the surface evolution exhibits strong corrections in the scaling
traditionally attributed to an intrinsic width~\cite{wolf,moro,evans,chavez}
that hampers the direct observation of the KPZ critical exponents in this model.
Continuous (coarse-grained) limits of the DB model in $d=1+1$ yield the KPZ equation to
leading order but inconsistencies were found in higher
dimensions~\cite{Katzav,Haselwandter}. Preceded by studies lying on finite-time  and -size
corrections~\cite{fabioBD,FabioPhysA2006} and  intrinsic width~\cite{wolf,moro,evans,chavez,tiago2},  a
direct observation of KPZ universality class for the BD model in
$d=1+1$ was obtained recently by means of thoroughgoing simulations of very
large systems and very long growth times~\cite{vvdensky,vvdensky2}. Recently, a
connection between the BD model and the KPZ class in 2+1 dimensions was possible
by unveiling the nature of the intrinsic width of the model~\cite{bd_box2d}.
It was shown that the leading contribution to the intrinsic width comes from the
short wavelength fluctuations in the height increments $\delta h$ along
the deposition events.
Besides, it was shown  that these effects can be suppressed using a
coarse-grained interface built from the original one~\cite{bd_box2d}; see
Sec.~\ref{model_methods}.

An important theoretical problem is the upper critical dimension $d_u$ above
which fluctuations become negligible. {Analytically, there is no
consensus on the value of $d_u$ (see discussions in Ref.~\cite{Pagnani13}) and
an appealing and recent non-perturbative renormalization group analysis rules
out $d_u=3+1$ but the approach losses reliability for $d\gtrsim3.5+1$ within
the approximations considered~\cite{Canet,Canet2}.} Moreover, numerical
simulations of models believed as belonging to the KPZ class practically discard
$d_u=4+1$~\cite{Pagnani13,Kim13,Schwartz2012,Perlsman2006,odor2010} and evidences up
$d_u=11+1$ have been recently reported~\cite{Kim2014,Rodrigues15,alves14} in
agreement with former
conjectures~\cite{Marinari2002,Tu1994,Ala-Nissila1993,Ala-Nissila1998}. While in
2+1 dimensions the generalization of the KPZ ansatz was supported by several
models~\cite{healyPRL,Oliveira13R,healyPRE}, its extension to $d>2$ was based on
numerical simulations~\cite{alves14} of the restricted-solid-on-solid (RSOS)
model~\cite{KK}.  In the present work, we investigate the BD model extending the
analysis of Ref.~\cite{bd_box2d} to $3+1$ and $4+1$ dimensions. We verify the
validity of the KPZ universality class, including exponents and its ansatz. We
also, revisited the values of the cumulants of $\chi$ presented in
Ref.~\cite{alves14} for RSOS model using  now more accurate estimates of
$\alpha$.

The paper is organized as follow. In the next section the model details and the
approach used are presented. In section \ref{results}, the results are
presented and discussed. The conclusions are summarized in section
\ref{conclusions}.

\section{Model and methods}
\label{model_methods}

The ballistic deposition growth model is implemented in $d+1$ hypercubic
lattices of size $L$ with periodic boundary conditions. The particles are
deposited one at a time at a randomly chosen position of a $d$-dimensional
substrate. Each particle is released perpendicularly to the substrate and
becomes permanently stuck at the first contact with either the deposit or
substrate~\cite{barabasi}. The original interface is defined as the highest
position of a particle at each site of the substrate. A time unity corresponds to
the aggregation of $L^d$ particles to the deposit. The simulations were carried
out on substrates of sizes up to $L=1024$ with averages over up to $N=2000$
independent samples in $d=3+1$. For  $d=4+1$, we consider systems of size up to
$L=228$ and up to $N=1000$ samples. The smaller the size the larger the number of
samples.

We also investigate surfaces using the prescription of Ref.~\cite{bd_box2d}. The
procedure consists in dividing the original surface in bins of lateral size
$\varepsilon$, the binning parameter, and using only the site of highest height
inside each bin to build a coarse-grained interface used to compute statistics.
The net effect is that the binned interface is smoother than the original one,
the latter characterized by many narrow and deep valleys. In $d=2+1$, it was
shown that the intrinsic width of the coarse-grained surfaces is strongly
reduced and, consequently, the strong corrections to the scaling fall
off~\cite{bd_box2d}. It was shown that the binning does not change the
non-universal constants $\Gamma$ and $\vinf$.

The non-universal  constants in the KPZ equation, Eq.~\eqref{eq:KPZ}, and in its
ansatz, Eq.~\eqref{eq:htcorr}, can be obtained using the approach hereafter
called Krug-Meakin (KM) method~\cite{krug90} that is described as follows.
From Eq.~\eqref{eq:htcorr}, the
asymptotic velocity is given by
\begin{equation}
\frac{d\lrangle{h}}{dt}=\vinf + \lrangle{g} t^{\beta-1}+\cdots,
\label{eq:dhdt}
\end{equation}
where $\lrangle{g}=\beta s_\lambda \Gamma^\beta \lrangle{\chi}$. So, plotting
$d\lrangle{h}/dt$ against $t^{\beta-1}$ renders a straight line for long times with
intercept providing $\vinf$ and the angular coefficient $\lrangle{g}$. The
latter plays an important role to determine the cumulant ratio
$R=\lranglec{\chi^2}/\lrangle{\chi}^2$, where $\lranglec{A^n}$ is the notation
for $n$th order cumulant of $A$; see subsection~\ref{sec:hds}. The parameter $\lambda$ is obtained by the
deposition on tilted large substrates with an overall slope $s$, for which a
simple dependence between asymptotic velocity and slope
\begin{equation}
v \simeq v_\infty+\frac{\lambda}{2}s^2
\label{eq:lambda}
\end{equation}
is expected for the KPZ equation~\cite{krug90}. We can use the relation~\cite{krug90}
\begin{eqnarray}
\Gamma=|\lambda| A^{1/\alpha},
\label{eq:Gamma}
\end{eqnarray}
where $\alpha$ is the roughness
exponent of the KPZ class, to determine the amplitude of the fluctuations. The
parameter $A$ is obtained from the asymptotic velocity $v_L$ of  finite systems
of size $L$~\cite{krug90} using the relation
\begin{equation}
 \Delta v = v_L-v_\infty \simeq -\frac{A\lambda}{2}L^{2\alpha-2}.
 \label{eq:vl}
\end{equation}

The KM analysis requires a prior accurate knowledge of the both growth and
roughness exponents. In $d=3+1$, we adopt the growth exponent
$\beta_{3+1}=0.184(5)$ reported by \'Odor \textit{et al.}~\cite{odor2010} since
it has a small uncertainty and was obtained for a model with small corrections
to the scaling using large systems of size $L=1024$. In $d=4+1$, we adopt the
recent estimate $\beta_{4+1}=0.158(6)$ determined by Kim and Kim ~\cite{Kim13}
using a RSOS model with an optimal height restriction parameter that improves
the corrections to the scaling. The determination of $\Gamma$ is extremely
sensitive to the value of the roughness exponent since it is used twice in the
analysis via Eqs.~\eqref{eq:Gamma} and \eqref{eq:vl}. In Ref.~\cite{alves14}, it
was used the exponents of \'Odor \textit{et al}.~\cite{odor2010}, that in
$d=4+1$ is $\alpha_{4+1}=0.245(5)$, and was found for RSOS model
$\Gamma^\mathrm{(Odor)}=240(50)$, that led to
$\lrangle{\chi}^\mathrm{(Odor)}_{4+1}=-1.00(5)$ and
$\lrangle{\chi^2}_\mathrm{c,4+1}^\mathrm{(Odor)}=0.09(1)$ (see
Ref.~\cite{alves14} or section~\ref{sec:hds} for the procedure to determine
these cumulants). Here, we revisit  the data of Ref.~\cite{alves14} using a more
recent estimate of Pagnani and Parisi~\cite{Pagnani13} given by
$\alpha_{4+1}=0.2537(8)$ that was obtained doing a thorough finite size analysis
and we find a different value $\Gamma^\mathrm{(Pagnani)}=105(8)$ that leads to
$\lrangle{\chi}^\mathrm{(Pagnani)}_{4+1}=-1.14(2)$ and
$\lrangle{\chi^2}_\mathrm{c,4+1}^\mathrm{(Pagnani)}=0.12(1)$ which are, in
absolute values, 14\% and 30\%, respectively, above the estimates of
Ref.~\cite{alves14}. Similarly, we revisit the data of Ref.~\cite{alves14} for
RSOS in $d=3+1$ using an former but with smaller uncertainties estimate of
$\alpha_{3+1}=0.3135(15)$ by Marinari \textit{et al}.~\cite{Marinari}, obtained
using the same method of Ref.~\cite{Pagnani13}, and we find
$\Gamma^\mathrm{(Marinari)}_{3+1}=15.8(6)$ in contrast with
$\Gamma^\mathrm{(Odor)}_{3+1}=38(3)$ that was found using
$\alpha_{3+1}=0.29(1)$~\cite{odor2010}. This difference  in $\Gamma$ leads to
first and second cumulants approximately 20\% and 50\% bigger than those found
using $\alpha_{3+1}=0.29(1)$. The results are summarized  in
table~\ref{tab:RSOS}.

\begin{table}[]
\centering
\caption{Non-universal parameter $\Gamma$ and cumulants of $\chi$ for RSOS with
height restriction parameter $m=2$ (data from Ref.~\cite{alves14}) obtained
using two different values of the roughness exponent  reported in the 
literature for each dimension.
}
\label{tab:RSOS}
\begin{tabular}{ccccc}
\hline\hline
$d$ & \multicolumn{2}{c}{3+1} & \multicolumn{2}{c}{4+1} \\\hline
Ref. &  ~~\'Odor~\cite{odor2010} & ~~Marinari~\cite{Marinari} & ~~\'Odor~\cite{odor2010}  & ~~Pagnani~\cite{Pagnani13}     \\\hline
$\alpha$            & 0.29(1) & 0.3135(15) & 0.245(5)  & 0.2537(8)      \\
$\Gamma$            & 38(3)   & 15.8(6)    & 240(50)    & 205(8)      \\
$\lrangle{\chi}$    & $-0.86$ & $-1.06$    & $-1.00$   & $-1.14$     \\
$\lranglec{\chi^2}$ & 0.12    & 0.18       & 0.09      & 0.12    \\\hline\hline   
\end{tabular}
\end{table}

\section{results and dicussions}
\label{results}

\subsection{Scaling and intrinsic width}
\label{sec:wint}
\begin{figure}[ht]
	\centering
	\includegraphics*[width=0.85\linewidth,angle=90]{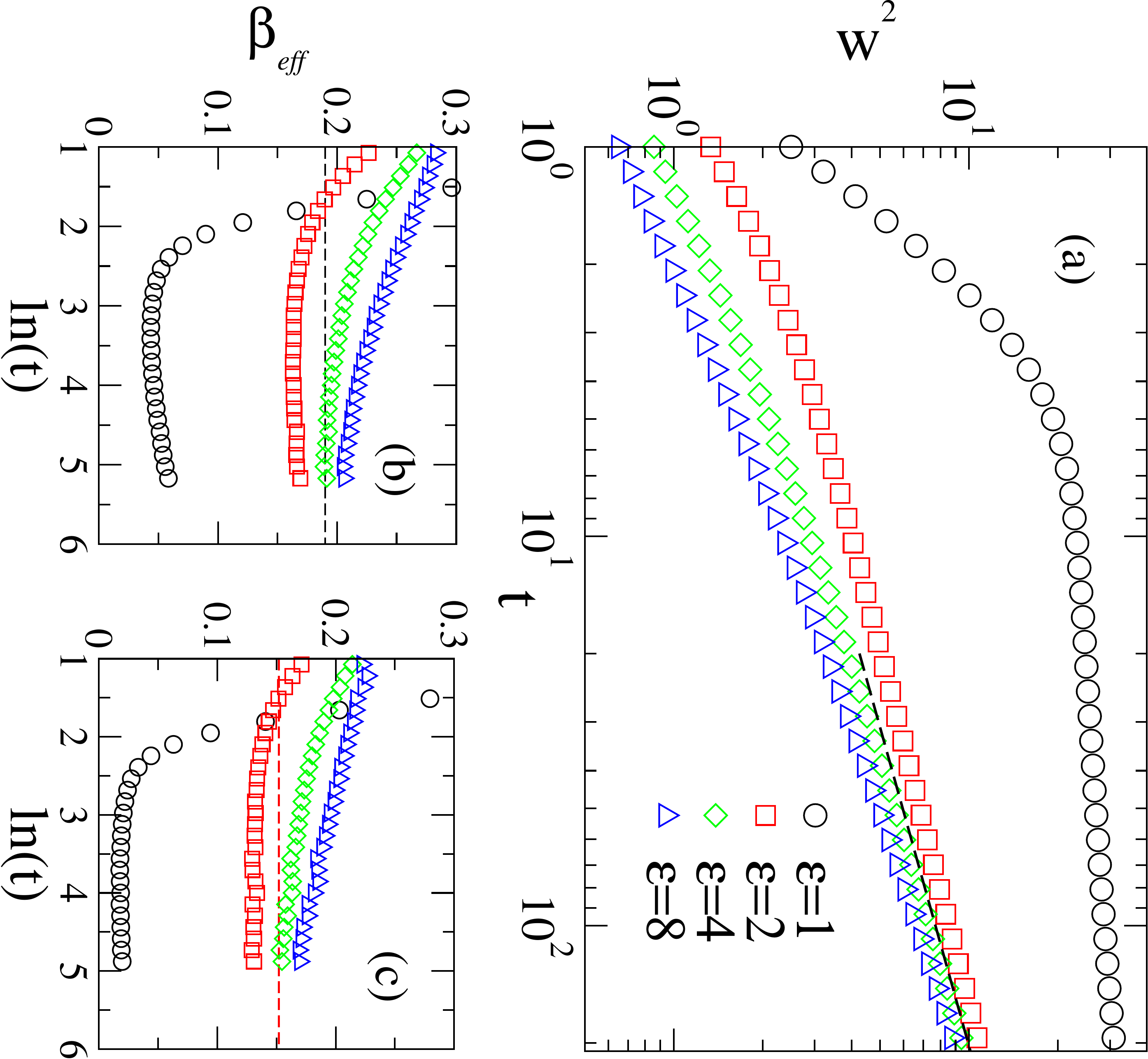}
	\caption{(a) Time evolution of the squared interface width of the BD model for
	both original ($\varepsilon=1$) and reconstructed surfaces  in $d=3+1$. The
	dashed line is a power law with exponent $2\beta = 0.368$. Similar behavior
	is observed for $d=4+1$. Effective growth exponents, $\beta_\mathrm{eff}=d(\ln
	w)/d(\ln t)$, are shown in bottom panels for (b) $d=3+1$ and (c) $4+1$. 	The
	dashed horizontal lines are the growth exponents found for other models in the
	KPZ class with small corrections to  the scaling in the respective
	dimensions~\cite{odor2010,Kim13}.}
	\label{fig:w2_ells}
\end{figure}
Figure~\ref{fig:w2_ells}(a) shows the interface width evolution in $d=3+1$
considering the original surface of the BD model as well as  those obtained with
binning parameters $\varepsilon=2$, 4 and 8. The effective growth exponents,
given by the local  derivative of $\ln w$ versus  $\ln t$, are shown in 
Figs.~\ref{fig:w2_ells}(b) and (c).  As aforementioned, the time evolution of the
interface width for the original BD surfaces exhibits strong corrections in the
scaling, leading to a very low effective  exponent $\beta_\mathrm{eff}$. In particular, $\beta_\mathrm{eff}$ becomes close to zero for
$d=4+1$ in the  investigated time interval, which is consistent with an upper
critical dimension $d_u=4$. However, as  in the previous $d=2+1$ 
analysis~\cite{bd_box2d}, a convergence to the KPZ growth exponent is observed
for the coarse-grained surfaces with $\varepsilon>1$ in both $d=3+1$ and 4+1,
see Fig.~\ref{fig:w2_ells}. Notice that there is an optimal interval of bin size
where the convergence becomes faster. Indeed, if the bin size is very small the
reconstructed surface still has  narrow and deep valleys  and thus a high
intrinsic width. On the other hand, if $\varepsilon$ is too large, only extremal
heights are accessed in the statistics and the convergence slows down.

\begin{figure}[ht]
	\centering
	\includegraphics*[width=0.85\linewidth]{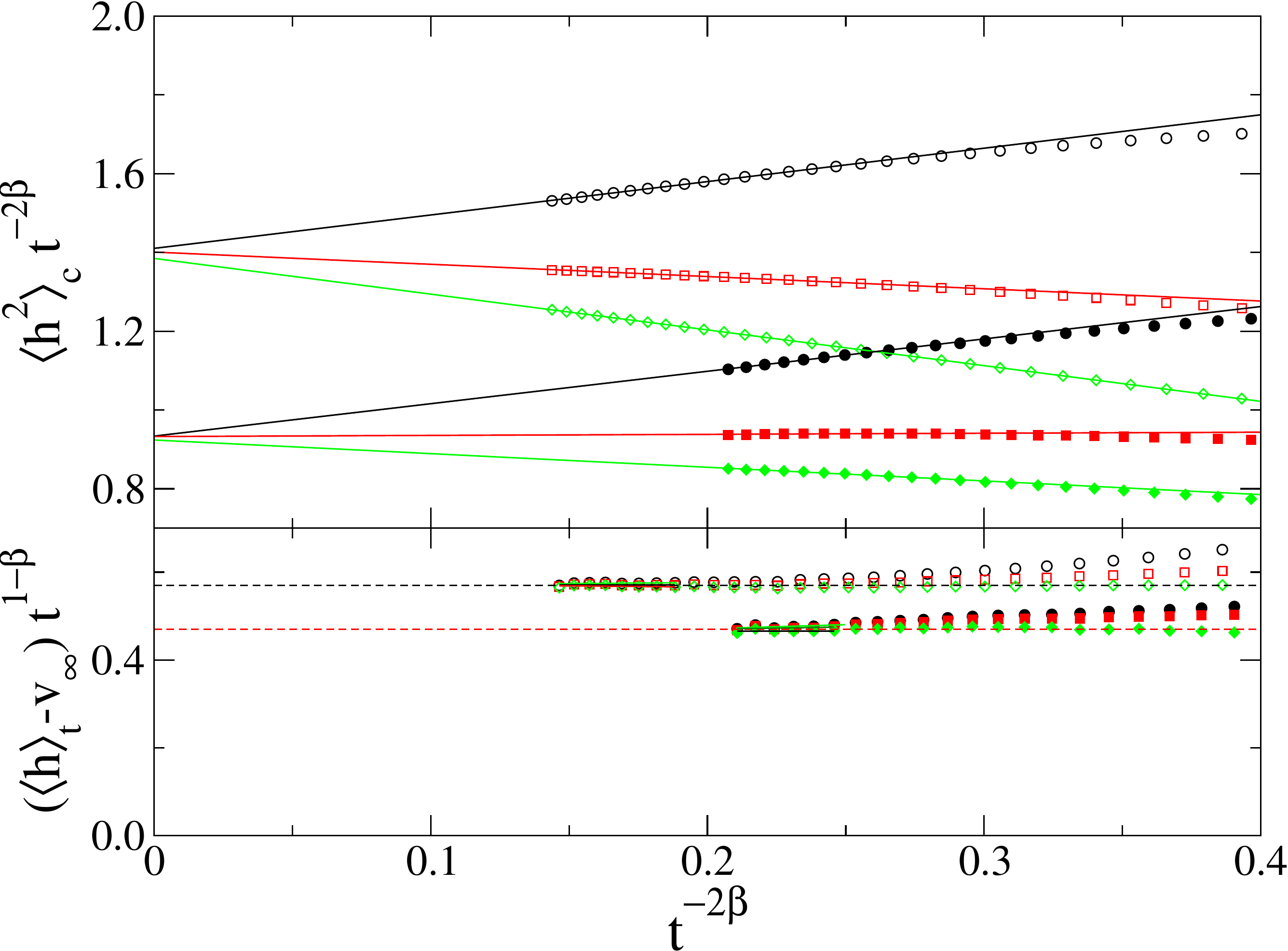}
	\caption{Determination of nonuniversal cumulants. Top:
	$\lranglec{g^2}=\Gamma^{2\beta} \lrangle{\chi^2}_c$  for $d=3+1$  (open
	symbols) and 4+1 (filled symbols)  for DB using binned substrates with
	$\varepsilon=2$, 4, and 8 from top to bottom. The lines are linear regressions
	used to determine $\lranglec{g^2}$. Bottom: determination of
	$\lrangle{g}$=$\beta\Gamma\lrangle{\chi}$ for $d=3+1$  (open symbols) and 4+1
	(filled symbols)  for DB using binned substrates with $\varepsilon=2$, 3, and
	4. Dashed lines are estimates of $\lrangle{g}$. 
} \label{fig:g2_e}
\end{figure}

The strong corrections observed in the interface width scaling can be reckoned
with an additive term, the squared intrinsic width
$w^2_i$~\cite{tiago2,evans,chavez,moro}, in the Family-Vicsek ansatz~\cite{FV}
as
\begin{equation}
 w^2(L,t) = L^{2\alpha}f\left(\frac{t}{L^{z}}\right) + w^2_i,
 \label{eq:FV}
\end{equation}
where  the scaling function $f(x)$ behaves as $f(x)\sim x^{2\beta}$ if $x\ll 1$
and $f(x)\sim \mbox{constant}$ if $x\gg 1$. The intrinsic width can be set in
terms of the KPZ ansatz, Eq.~\eqref{eq:htcorr}, as~\cite{bd_box2d}
\begin{equation}
w^2_i = \langle h^2\rangle_c - (\Gamma t)^{2\beta} \langle\chi^2\rangle_c.
\label{eq:wi}
\end{equation}
According to Eq.~\eqref{eq:htcorr}, the second cumulant of the height is given by
\begin{equation}
\label{eq:h2cum}
\langle h^2\rangle_c = (\Gamma t)^{2\beta} \langle\chi^2 \rangle_c +
 2(\Gamma t)^{\beta} \mathrm{cov}(\chi,\eta) + \langle\eta^2\rangle_c + \ldots,
\end{equation}
where
$\mathrm{cov}(\chi,\eta)=\langle\chi\eta\rangle-\langle\chi\rangle\langle\eta\rangle$. The  cumulant $\lranglec{g^2}=\Gamma^{2\beta} \lrangle{\chi^2}_c$~\cite{Alves13}, necessary to compute $w_i$, can be estimated considering the long time limit of
\begin{equation}
\lranglec{g^2} = \lim_{t \to \infty}\frac{\langle h^2\rangle_c}{t^{2\beta}}
\end{equation}
Assuming that there is no statistical dependence between $\chi$ and $\eta$,
$\mathrm{cov}(\chi,\eta)=0$, a linear extrapolation to $\lranglec{g^2}$ is
expected in curves $\langle h^2\rangle_c/t^{2\beta}$ against $t^{-2\beta}$, as
confirmed in Fig.~\ref{fig:g2_e} in both dimensions for three values of the
binning parameter. Propagating the uncertainties in the growth 
exponents, the estimated values are
$\lranglec{g^2}^{(3+1)}=1.4(1)$ and  $\lranglec{g^2}^{(4+1)}=0.93(8)$; see table~\ref{tab:parBD}.

\begin{figure*}[ht]
	\centering
	\includegraphics[width=0.33\linewidth]{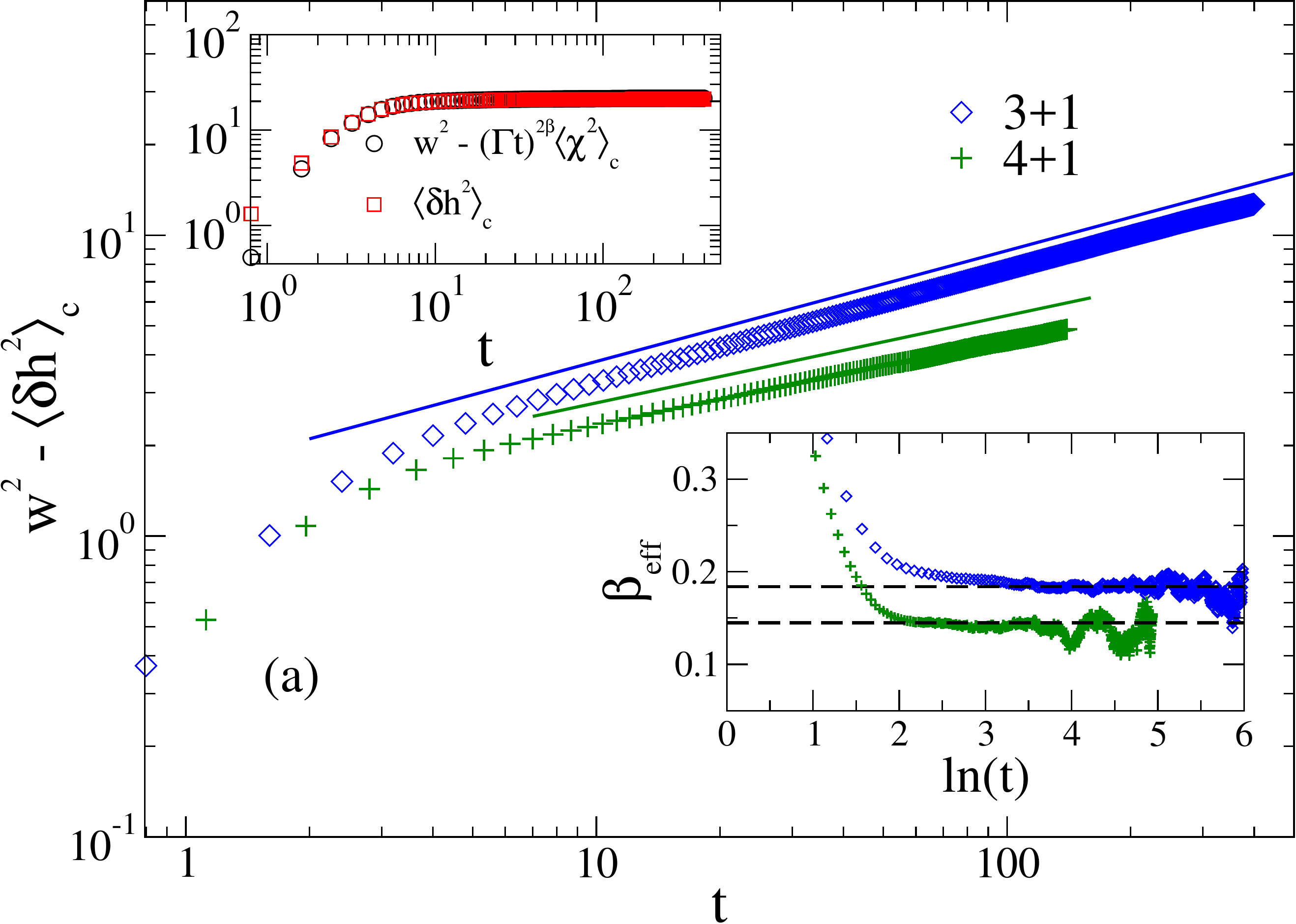}
	\includegraphics[width=0.33\linewidth]{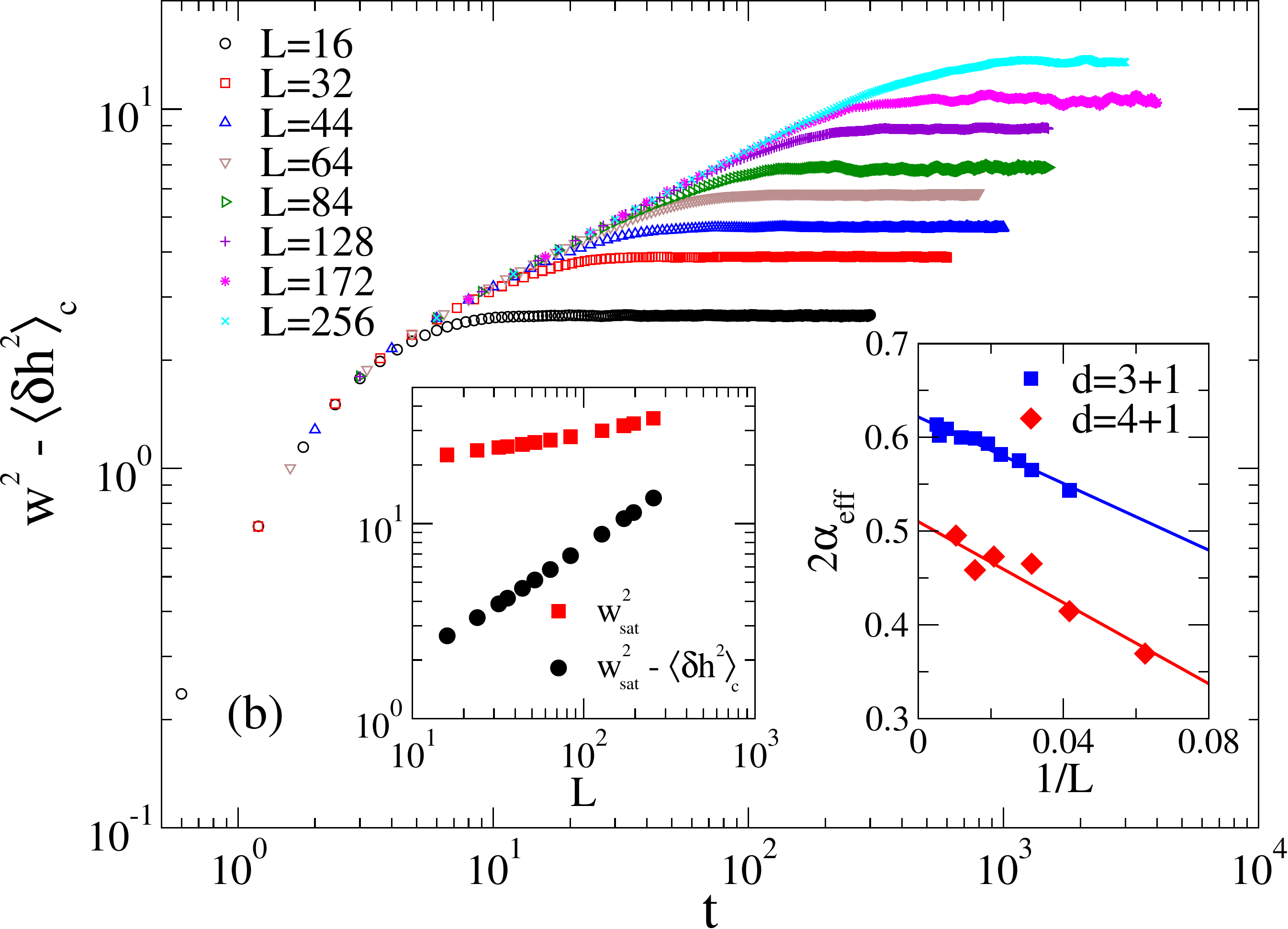}
	\includegraphics[width=0.31\linewidth]{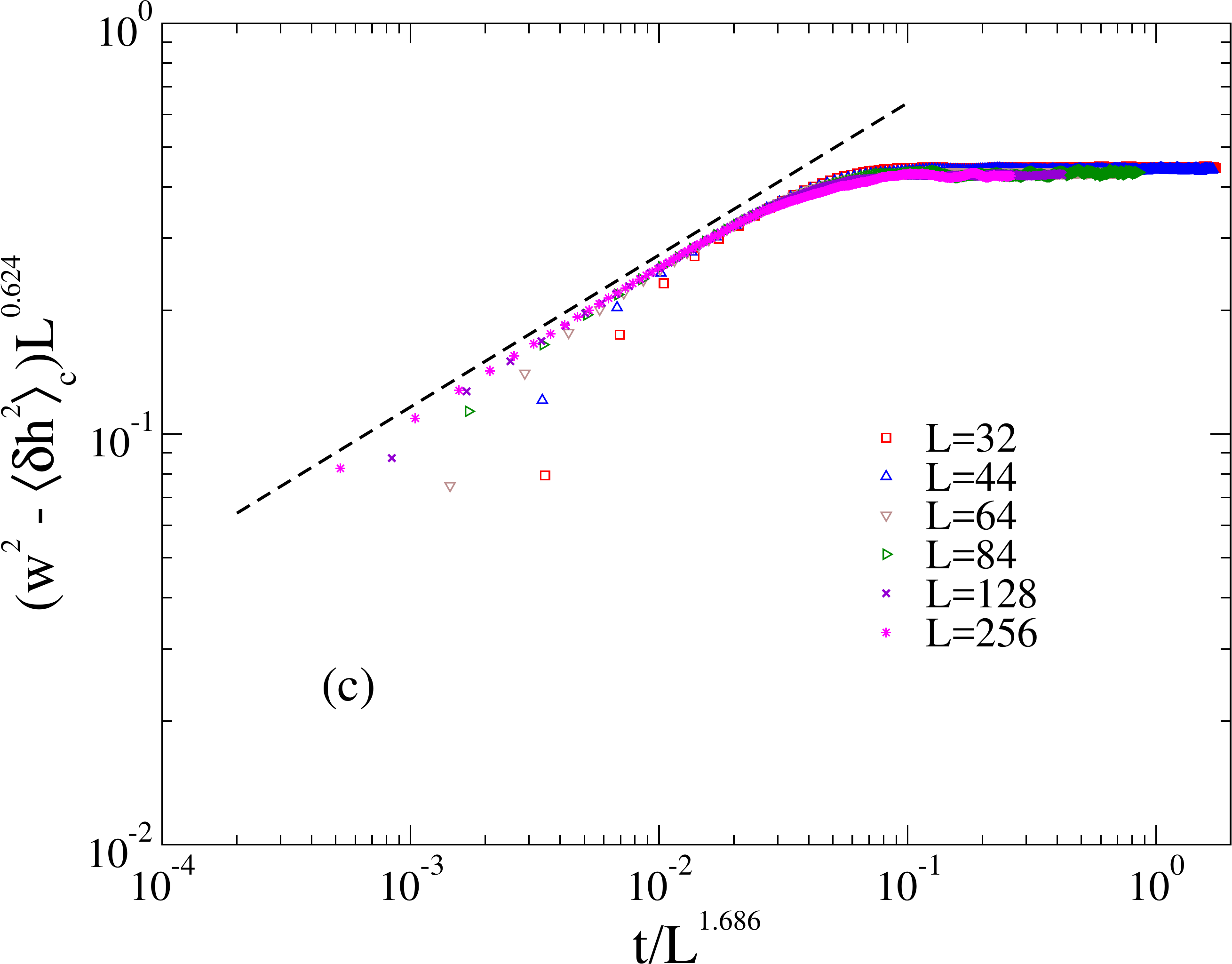}
	\caption{\label{fig:w2dh2} Interface width analysis for BD in $d=3+1$
	and $4+1$. (a) Main panel: squared interface width discounting the second
	cumulants of height increments for systems of sizes $L=1024$ and $228$ in
	$d=3+1$ and 4+1, respectively. The  lines are power laws with exponents $2\beta
	= 0.368$ and $0.290$. Top inset: time evolution of the intrinsic width and
	second cumulant of $\delta h$ for $d=3+1$. Bottom inset:
	effective growth exponent analysis. (b) Main panel:  squared interface width
	discounting the second cumulants of height increments for $d=3+1$ and different
	sizes. Left inset: Saturated squared interface width discounting or not 
	the second cumulant of $\delta h$ against lattice size for $d=3+1$.
	Right inset: Effective roughness exponent analysis for $d=3+1$ and 4+1. (c) Squared
	interface width in $d=3+1$ scaled with the exponents found in our analysis.}
\end{figure*}

The leading contribution to the intrinsic width in $d=2+1$ comes from the large
fluctuations of the height increments in the deep valleys of the BD
interfaces~\cite{bd_box2d}: $w^2_i\approx \langle (\delta h)^2\rangle_c$ where
$\delta h(i,t) =h(i,t+dt)-h(i,t)$ is the increment at site $i$ at a step time
$dt=1/L^d$. In the present work, we verify that this conjecture is still
accurate for $d=3+1$ and 4+1. The upper inset of Fig.~\ref{fig:w2dh2}(a) shows
the time evolution of the squared intrinsic width, Eq.~\eqref{eq:wi}, and of the
second cumulant of $\delta h$. We observe a very good accordance between these
quantities. The intrinsic widths found for long times, propagating the
uncertainties in both $\beta$  and $\lranglec{g^2}$, were $w_i^{(3+1)}=21.1(1)$
and $w_i^{(4+1)}=32.6(1)$ while for the height increments we found
$\lranglec{(\delta h)^2}=21.13$ and 32.10 in $d=3+1$ and 4+1, respectively; see table~\ref{tab:parBD}. This
shows that the corrections in the scaling become more relevant at higher
dimensions and explains why it is currently impossible to see KPZ exponents in
the high dimensional BD model using a plain analysis. We also compared the third
cumulant of $\delta h$ with $\langle h^3\rangle_c - (\Gamma t)^{3\beta}
\langle\chi^3\rangle_c$ and a small but relevant difference was found, as in
$d=2+1$~\cite{bd_box2d}, showing a non-trivial relation between the height
increments and corrections terms in Eq.~\eqref{eq:htcorr}. 

\begin{table}[]
	\centering
	\caption{Non-universal parameters for BD model. }
	\label{tab:parBD}
	\begin{tabular}{ccccc}
		\hline\hline
		$d$ & $\lrangle{g}$& $\lranglec{g^2}$ & $\lranglec{(\delta h)^2}$ &  $w_i^2$ \\\hline
		3+1 & 0.568(5) & 1.40(1) & 21.13 & 21.1(1)      \\
		4+1 & 0.466(5) & 0.93(8)  & 31.10 & 32.6(1)    \\\hline\hline   
	\end{tabular}
\end{table}

The evolution of the interface width discounting $\langle (\delta h)^2\rangle_c$
for original BD interfaces is shown in the main panel of the
Fig.~\ref{fig:w2dh2}(a). Differently from the binning procedure, this method is
free from adjustable parameters. The growth exponents found were
$\beta_{3+1}=0.185(5)$ and $\beta_{4+1}=0.145(10)$, in sharp agreement with the
exponent $\beta_{3+1}=0.184(5)$ of Ref.~\cite{odor2010} and in marginal
agreement with the recent estimate $\beta_{4+1}=0.158(6)$ of
Ref.~\cite{Kim2014}, as can be seen in the effective exponent analysis in the
bottom inset of Fig.~\ref{fig:w2dh2}(a). Here, it is worth to note that the
intrinsic width is slightly larger than $\lrangle{(\delta h)^2}_c$ in $d=4+1$,
that, together with the finite time used, can  explain the slightly smaller
growth exponent found in this dimension. This strategy can be used to obtain the
roughness exponent $\alpha$ as well. The squared interface width discounting
$\lrangle{(\delta h)^2}_c$ is shown as a function of time for different sizes and
$d=3+1$ in Fig.~\ref{fig:w2dh2}(b). The left inset compares the saturated values of $w^2$  and
$w^2-\lrangle{(\delta h)^2}_c$. We see that the intrinsic width is still much
larger than the long wavelength interface width, obtained discounting the
intrinsic one, even for the largest investigated size of $L=256$. Note that
$\lranglec{(\delta h)^2}$ has as small but not negligible dependence with size
that was reckoned in our analysis. The right inset of Fig.~\ref{fig:w2dh2} shows the effective roughness
exponent analysis for $d=3+1$ and 4+1. The estimated values of roughness
exponents are $\alpha_{3+1}=0.312(2)$ and $\alpha_{4+1}=0.251(5)$ that, withing
uncertainties, agree very well with the both estimates
$\alpha_{3+1}^\mathrm{(Marinari)}=0.3135(15)$~\cite{Marinari}  and
$\alpha_{4+1}^\mathrm{(Pagnani)}=0.2537(8)$~\cite{Pagnani13}. Considering our
estimates for the growth exponent we found $\alpha+\alpha/\beta=2.00(15)$ and
1.98(15) in $d=3+1$ and 4+1 dimensions, respectively, in agreement with Galilean
invariance scaling relation~\cite{KPZ}. In Fig.~\ref{fig:w2dh2}(c), we confirm
the validity of the modified Family-Vicsek ansatz, Eq.~\eqref{eq:FV},  showing
the collapse of $(w^2-\lrangle{(\delta h)^2}_c)/L^\alpha$ against
$t/L^{\alpha/\beta}$ in $d=3+1$ for different systems sizes onto a universal
curve.

\subsection{Height Distribution}
\label{sec:hds}

\begin{figure}[ht]
	\centering
\includegraphics*[width=0.9\linewidth]{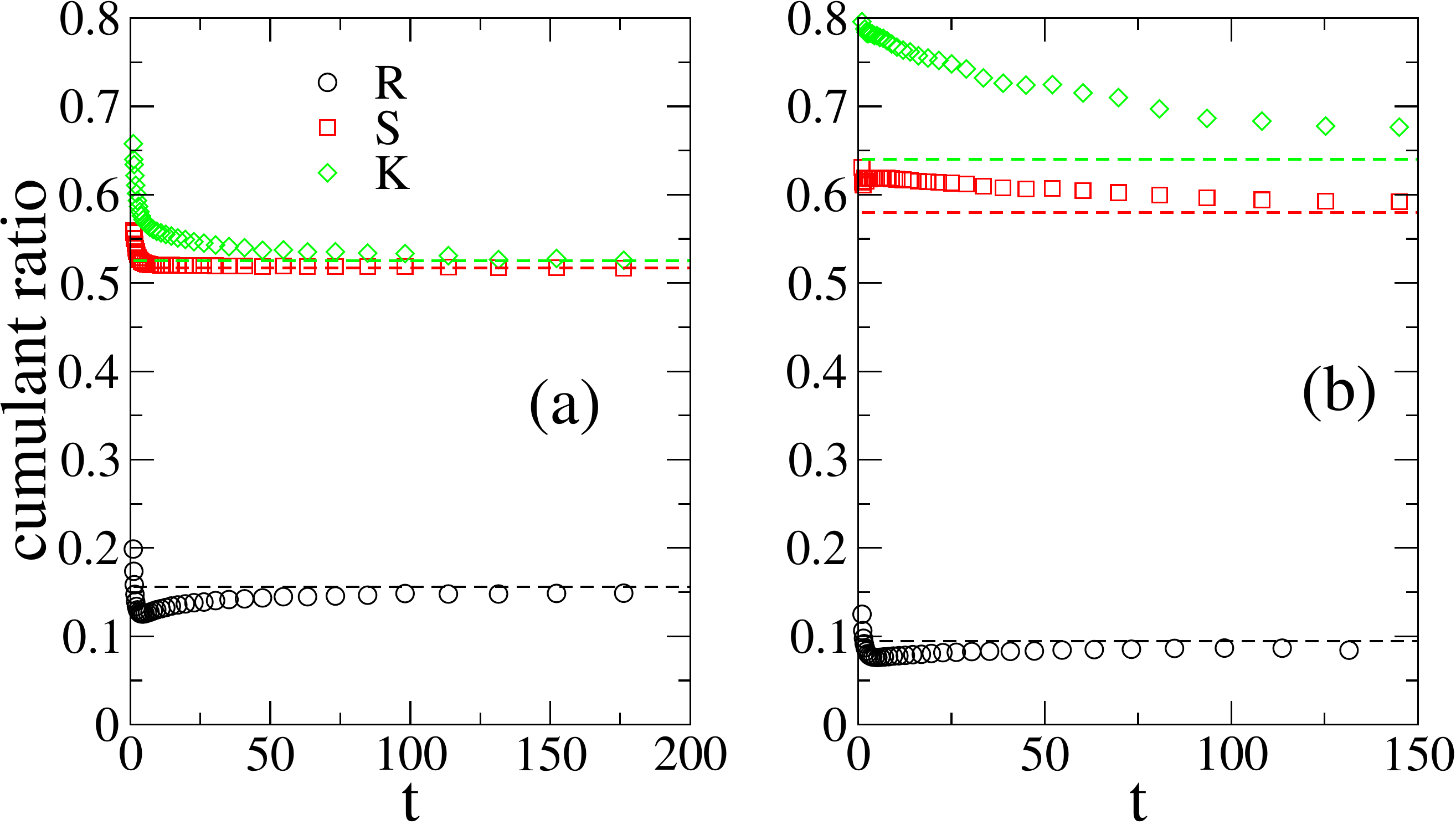}
\caption{Determination of dimensionless cumulant ratios for BD in (a) $d=3+1$
and (b) $4+1$. Dashed lines represent the estimates of cumulant ratios for RSOS
model taken from Ref.~\cite{alves14}. The BD results were obtained using a binning
parameter $\varepsilon=4$.}
	\label{fig:RSK}
\end{figure}

Let us now focus on the random variable $\chi$ of the  KPZ ansatz. An initial
assessment involves dimensionless cumulant ratios which can be determined without
knowing the constants $\Gamma$ and $\vinf$. The skewness $S$ and kurtosis $K$
are given by
\begin{equation}
S=\frac{\lranglec{\chi^3}}{\lranglec{\chi^2}^{1.5}}=\lim\limits_{t\rightarrow\infty} \frac{s_\lambda\lranglec{h^3}}{\lranglec{h^2}^{1.5}}
\end{equation}
and
\begin{equation}
K=\frac{\lranglec{\chi^4}}{\lranglec{\chi^2}^{2}}=\lim\limits_{t\rightarrow\infty} \frac{\lranglec{h^4}}{\lranglec{h^2}^{2}},
\end{equation}
being the right-hand sides obtained with Eq.~\eqref{eq:htcorr}. Another useful cumulant ratio is
given  by~\cite{Oliveira13R,Alves13}
\begin{equation}
	R=\frac{\lranglec{\chi^2}}{\lrangle{\chi}^{2}} = \frac{\beta^2\lranglec{g^2}}{\lrangle{g}^{2}},
\end{equation}
were $\lrangle{g}=\beta\Gamma^\beta\lrangle{\chi} =
\lim\limits_{t\rightarrow\infty}(\lrangle{h}_t-\vinf)t^{1-\beta}$, see
Eq.~\eqref{eq:dhdt} and Fig.~\ref{fig:g2_e}. The analyses of these cumulant are
shown in Fig.~\ref{fig:RSK}. We can see that the cumulant ratios are either very
close or approaching the  values obtained for the RSOS
model\footnote{Differently from the cumulants of $\chi$ (see
	section~\ref{model_methods}), the cumulant ratios obtained for RSOS model in
	Ref.~\cite{alves14} are reliable references because the model has small
	finite-time corrections and the determination does not depend on $\alpha$.}in Ref.~\cite{alves14},
 corroborating these KPZ signatures for BD in higher
dimensions.

\begin{figure}[ht]
	\centering
	\includegraphics*[width=0.75\linewidth]{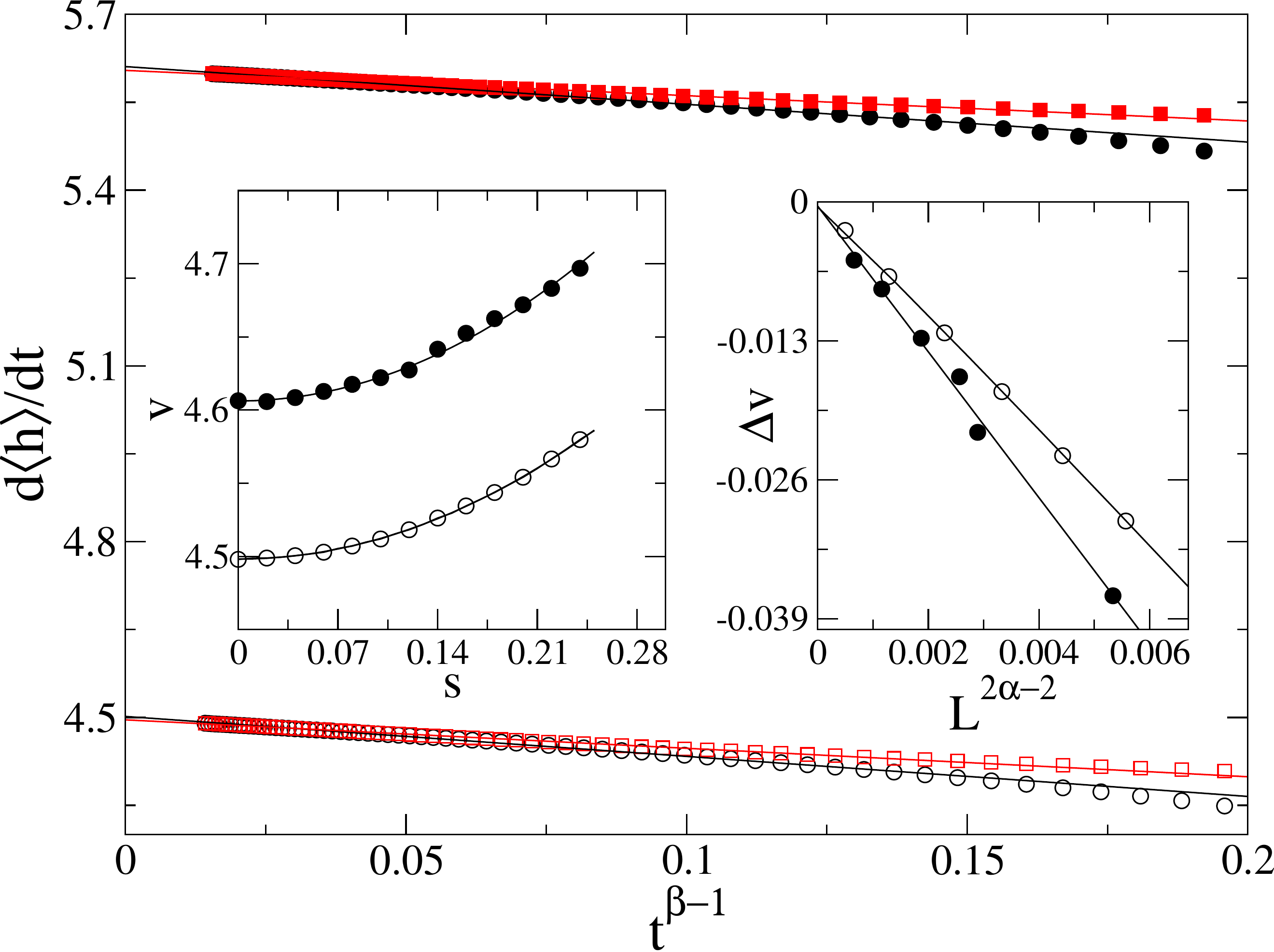}
	\caption{Parameter determination using KM method~\cite{krug90}. Main panel:
	Interface growth velocity for BD in $d=3+1$ (bottom curves) and 4+1 (top
	curves) are represented  by open and filled  symbols, respectively. We show the
	results for the original surface (squares) and binning parameter
	$\varepsilon=4$ (circles). Left Inset: growth velocity against substrate slope
	for $d=3+1$ (open symbol) and 4+1 (filled symbols). The velocity in $d=4+1$ is
	subtracted by 1 to improve visualization. Right inset: linear dependence of
	the velocity difference $\Delta v=v_L-\vinf$  with the system size according
	Eq.~\eqref{eq:vl}.}
	\label{fig:vinf}
\end{figure}

To numerically determine the probability distribution function  $\rho(\chi)$ requires
accurate estimates of the non universal constants $\vinf$ and $\Gamma$. The
determination of the asymptotic velocities for $d=3+1$ and 4+1 are shown in the
main panel of the Fig.~\ref{fig:vinf}. As observed in $d=2+1$~\cite{bd_box2d},
the asymptotic growth velocity is independent of $\varepsilon$, and converges to
the same value as the original surface. Our estimated values of the velocity are
$v_{\infty,3+1}=4.49820(2)$ and $v_{\infty,4+1}=5.60615(5)$, see
table~\ref{tab:KMBD}. Notice that since the asymptotic velocity does not 
dependent on $\varepsilon$, the KM analysis also does not. The determination of
$\lambda$ using  Eq.~\eqref{eq:lambda},  shown in the left inset of
Fig.~\ref{fig:vinf}, provides $\lambda_{3+1}=2.81(1)$ and
$\lambda_{4+1}=3.17(4)$.

The KM curves used to determine the values of $\lambda A$,
Eq~\eqref{eq:vl}, with the roughness exponents
$\alpha_{4+1}^\mathrm{(Marinari)}=0.3135(15)$ and
$\alpha_{4+1}^\mathrm{(Pagnani)}=0.2537(8)$, are shown in the right inset of
Fig.~\ref{fig:vinf}. The values of $\Gamma=\vert \lambda \vert A^{1/\alpha}$
found are $\Gamma_{3+1}^\mathrm{(Marinari)}=205(20)$ and
$\Gamma_{4+1}^\mathrm{(Pagnani)}=730(30)$. Using our exponents,
$\alpha_{3+1}=0.312(2)$ and $\alpha_{4+1}=0.251(5)$ we have found
$\Gamma_{3+1}=215(15)$ and $\Gamma_{4+1}=700(200)$. Using the exponent of
Ref.~\cite{odor2010}, $\alpha_{3+1}=0.29(1)$ and $\alpha_{4+1}=0.245(5)$, we have found
$\Gamma_{3+1}^\mathrm{(Odor)}=500(200)$ and
$\Gamma_{4+1}^\mathrm{(Odor)}=1200(250)$, both presenting large uncertainties
and in odds with the previous estimates. In the remaining of the analysis we use
$\Gamma_{3+1}=205(20)$ and $\Gamma_{4+1}=730(30)$ remarking that using the
estimates with the exponent of Ref.~\cite{odor2010} leads to values consistent
with our previous analysis of Ref.~\cite{alves14}. The KM parameters are summarized
in Table~\ref{tab:KMBD}.

\begin{table}[]
	\centering
	\caption{Non-universal KM parameters and cumulants of $\chi$ for DB model.}
	\label{tab:KMBD}
	\begin{tabular}{cccccc}
		\hline\hline
	$d$& $\vinf$    & $\lambda$ & $\Gamma$  & $\lrangle{\chi}$ & $\lranglec{\chi^2}$  \\\hline
	3+1& 4.49820(2) & 2.81(1)     & $205(20)$ & -1.15(3)          & 0.197(7)              \\
	4+1& 5.60615(5) & 3.17(4)      & $730(30)$ & -1.04(1)          & 0.115(3)          \\ \hline\hline   
	\end{tabular}
\end{table}

Possessing the KM parameters, the first and second cumulants of $\chi$  can be obtained 
directly from 
\begin{equation}
\lrangle{\chi}=\frac{\lrangle{g}}{\beta \Gamma^\beta}
\end{equation}
and
\begin{equation}
\lranglec{\chi^2}=\frac{\lranglec{g^2}}{\Gamma^{2\beta}},
\end{equation}
where $\lrangle{g}$ and $\lranglec{g^2}$ are defined in Sec.~\ref{model_methods}
and shown in table~\ref{tab:KMBD}. The results are $\lrangle{\chi}_{3+1}=1.15(3)$,
$\lrangle{\chi}_{4+1}=1.04(1)$, $\lrangle{\chi^2}_{c,3+1}=0.197(7)$ and 
$\lrangle{\chi^2}_{c,4+1}=0.115(3)$, which are in very good agreement with the corresponding
cumulants for RSOS shown in table~\ref{tab:parBD}. These cumulants are summarized in 
table~\ref{tab:KMBD}.

\begin{figure}[th]
\centering
\includegraphics[width=0.9\linewidth]{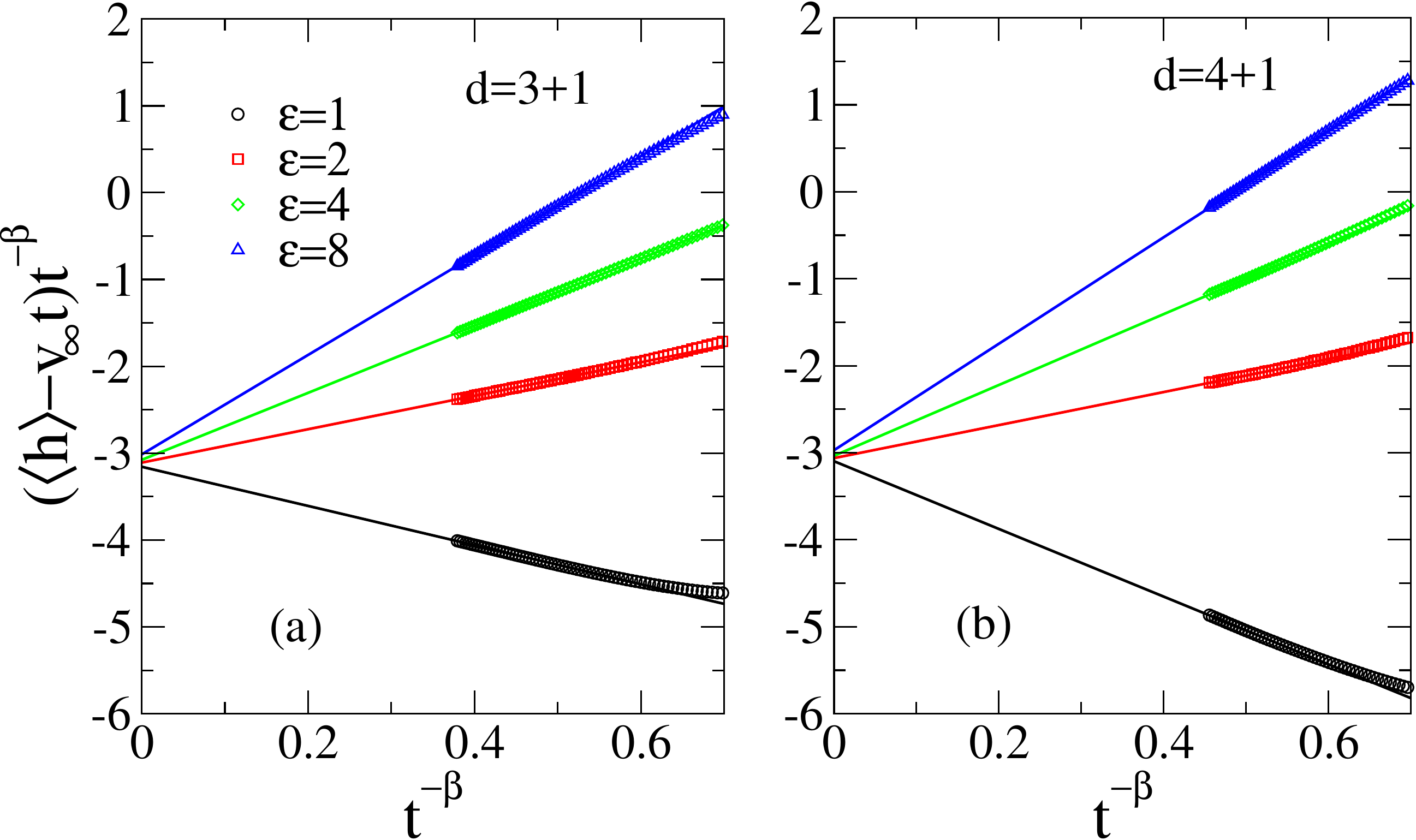}
\caption{Determination of the average shift $\lrangle{\eta}$ in (a) $d=3+1$ and (b) 4+1.}
\label{fig:eta3e4}
\end{figure}

Lets us define the random variable
\begin{equation}
q'=\frac{h-\vinf t-\lrangle{\eta}}{(\Gamma t)^\beta}
\label{eq:qprime2}
\end{equation}
whose probability  distribution function converges to $\rho(\chi)$ as $t\rightarrow\infty$~\cite{Alves13,Oliveira13R}.
To determine the parameter $\lrangle{\eta}$ we use that 
\begin{equation}
\frac{\lrangle{h}-\vinf t}{t^\beta}=\Gamma^\beta \lrangle{\chi}+\lrangle{\eta}t^{-\beta}+\cdots,
\end{equation} 
such that plotting this left-hand quantity against $t^{-\beta}$ extrapolates linearly to $\Gamma^\beta \lrangle{\chi}$
and the angular coefficient is $\lrangle{\eta}$. Figure~\ref{fig:eta3e4} confirms the expected behavior
and the existence of the correction. Since $\eta$ is a short wavelength correction, 
the value of $\lrangle{\eta}$ depends on the binning parameter $\varepsilon$~\cite{bd_box2d},
as shown in table~\ref{tab:eta}.

\begin{table}[]
	\centering
	\caption{Average value of the correction $\eta$ in the KPZ ansatz,
		Eq.~\eqref{eq:htcorr} and Fig.~\ref{fig:eta3e4}.}
	\label{tab:eta}
	\begin{tabular}{ccccc}
		\hline\hline
		$d$ & $\epsilon=1$ ~ & $\epsilon=2$ ~ & $\epsilon=4$ ~ & $\epsilon=8$ \\\hline
		3+1 & -2.3 & 1.9 & 3.9 & 5.7      \\
		4+1 & -3.9 & 1.9 & 4.1 & 6.1    \\\hline\hline   
	\end{tabular}
\end{table}

\begin{figure}[ht]
\centering
\includegraphics*[width=7cm]{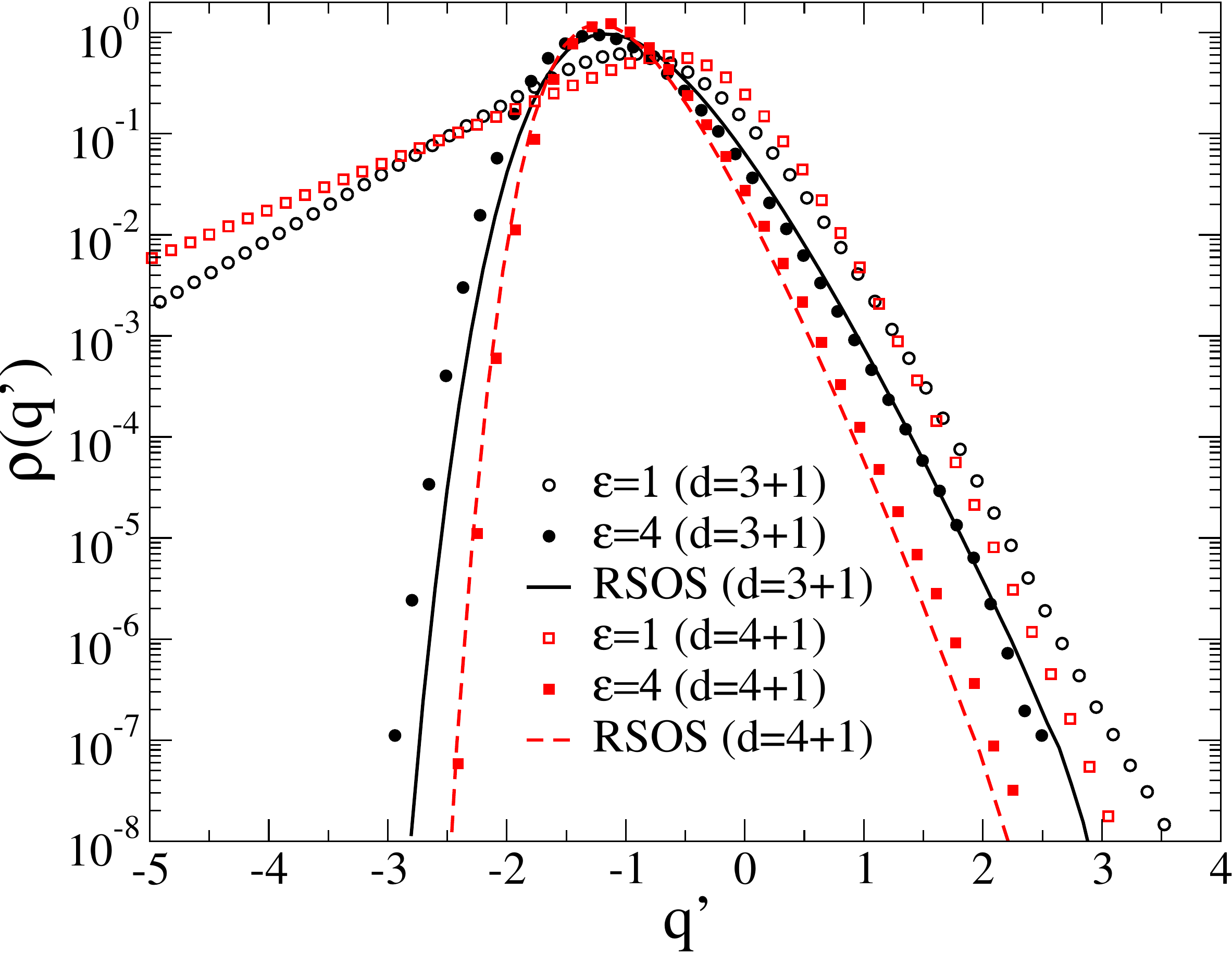}
\caption{Comparison of the probability distribution function,
Eq.~\eqref{eq:qprime2}, of the original and binned surfaces of the BD model in
$d=3+1$ and 4+1 dimensions with the RSOS model. The growth times in BD models
are $t=190$ and 145 for $d=3+1$ and 4+1, respectively.}
\label{fig:pchi_ells}
\end{figure}

In Fig.~\ref{fig:pchi_ells}, the probability distribution functions for binned
surfaces in $d=3+1$ and 4+1 are compared with those of the original interface as
well as with those of  RSOS model, the last one built using the estimates of
$\Gamma$ of  table~\ref{tab:RSOS} while the other parameters are those reported in
Ref.~\cite{alves14}. If, on the one hand, the original surfaces are not close to the RSOS
distributions, on the other hand, we see a satisfactory agreement with the binned surfaces,
presenting small deviations in either left or right tails for $d=3+1$ and 4+1,
respectively. These deviations must shrink if much longer growth
times are considered.

\section{Conclusions}
\label{conclusions}

Ballistic deposition growth models are characterized by a prominent lateral
growth and therefore are considered standards of KPZ growth~\cite{barabasi}.
However, strong finite-time and -size corrections make a direct realization of
the KPZ exponents in higher dimensions extremely hard and, in practice,
unaccessible with our current computer resources. However, eliciting the origin
of the leading contributions to the corrections as being due to the fluctuations
of height increments along the deposition of particles,  it was possible to do a
connection between ballistic deposition and KPZ universality class in $d=2+1$
dimensions~\cite{bd_box2d}. Moreover, using the coarse-grained surface where
only the highest points inside small bins of size $\epsilon\ll \xi$, where $\xi$
is the surface correlation length, it was possible to obtain the KPZ exponents
as well as the universal underlying stochastic fluctuations of the KPZ class in
$d=2+1$~\cite{bd_box2d}.

In the present work, we show that the methodology of Ref.~\cite{bd_box2d}
remains valid for ballistic deposition in $d=3+1$ and 4+1 dimensions. We observe
that the squared intrinsic width is given by $w_i^2\approx \lrangle{(\delta
	h)}_c$, where $\delta h$ is the height increment in a deposition step, and
becomes more relevant at higher dimensions. Growth and roughness exponents in
very good agreement with those reported for KPZ models with small corrections to
the scaling~\cite{odor2010,Kim13,Pagnani13,Marinari2002,alves14} were obtained
when the intrinsic width was explicitly reckoned in the scaling analysis. Using a
binned surface analysis, we also provide evidences that the underlying
fluctuation $\chi$ of height profiles belongs to the KPZ class, using the
dimensionless cumulant ratios and the  probability distribution function itself.

We also revisit the data for RSOS deposition model  reported in Ref.
\cite{alves14} considering more accurate estimates of the roughness exponents.
We have found that the non-universal parameter $\Gamma$, representing the
amplitude of the interface fluctuations, changes significantly  implying in
changes of the estimates of the cumulants of $\chi$.

Finally, it is worth noticing that our results provide a new numerical evidence
for an upper critical dimension, if it exists, larger than $d=4+1$ corroborating
former~\cite{Tu1994,Ala-Nissila1993,Ala-Nissila1998,Marinari2002} and
recent~\cite{Pagnani13,Kim2014,alves14,Rodrigues15} findings.

\begin{acknowledgments}
Authors acknowledge the support from CNPq and FAPEMIG (Brazilian agencies).
\end{acknowledgments}

%

\end{document}